\documentclass[letterpaper]{sig-alternate-10pt}
\usepackage{times}
\usepackage{algorithm,algorithmic,amsmath,endnotes,epsfig,graphicx,subcaption,color,balance,pifont,multirow}
\usepackage{enumitem}
\usepackage{multirow}
\setlist{nosep}

\newcommand{\xmark}{\ding{55}}%

\newcommand{\headline}[1]{\vspace{+0.03in}\noindent\textbf{#1}}

\newcommand{\symnet}{SymNet }
\begin{document}

\author{Radu Stoenescu, Matei Popovici, Lorina Negreanu, Costin Raiciu\\
  University Politehnica of Bucharest}

\title{SymNet: scalable symbolic execution for modern networks}
\maketitle


\newcommand{\mateisize}{\scriptsize}

\section*{Abstract}
We present SymNet, a network static analysis tool based on symbolic execution.
\symnet quickly analyzes networks by injecting symbolic 
packets and tracing their path through the network. 
Our key novelty is SEFL, a language we designed 
for network processing that is symbolic-execution friendly. 

\symnet is easy to use: we have developed parsers that automatically
generate SEFL models from router and switch tables, firewall
configurations and arbitrary Click modular router configurations.
Most of our models are exact and have optimal branching factor.
Finally, we built a testing tool that checks SEFL models 
conform to the real implementation.
\symnet can check networks containing routers with hundreds of thousands of prefixes 
and NATs in seconds, while ensuring packet header memory-safety and capturing network 
functionality such as dynamic tunneling, stateful processing and encryption. 
We used \symnet  to debug middlebox interactions
documented in the literature, to check our department's network and the Stanford
backbone network. Results show that symbolic execution is fast and more accurate than existing static analysis tools.
\section{Introduction}
\label{sec:introduction}

Modern networks deploy a mix of traditional switches and routers
alongside more complex network functions including security
appliances, NATs and tunnel endpoints. Understanding end-to-end
properties such as TCP reachability is difficult before
deploying the network configuration, and deployment can disrupt live
traffic. Static analysis of network dataplanes allows cheap, fast and 
exhaustive verification of deployed networks for packet reachability,
absence of loops, bidirectional forwarding, etc.  
All static analysis tools take as input a \emph{model} 
of each network box processing, the links
between boxes and a snapshot of the forwarding state, and are able to answer queries
about the network without resorting to dynamic testing \cite{staticchecking,hsa,nod,anteater,berk}. 

What is the best modeling language for networks? 
If possible, we
should simply use the implementation of network boxes (e.g. a C program), as this is
the most accurate and is easiest to use. If we view packets as
variables being passed between different network boxes, static network
analysis becomes akin to software testing. This is a problem that has
been studied for decades, and the leading approach is to use \emph{symbolic execution} \cite{klee}. 

Symbolic execution is really powerful: it explores all possible paths
through the program, providing possible values for each (symbolic)
variable at every point. In the context of static network
analysis, the power of symbolic execution lies in its ability to
relate the outgoing packets to the incoming ones: even if all the
incoming packet headers are unknown, a symbolic execution engine can
detect which header fields are allowed in each part of the network, 
which ones are invariant, and can tell \emph{how} the modified headers depend on the input when
they are changed. Unfortunately, symbolic execution scales poorly: its
complexity is roughly exponential in the number of branching
instructions (e.g. ``if'' conditionals) in the analyzed
program. Applying symbolic execution to actual network code quickly
leads to untenable execution times, as shown in
\cite{symbex-dobrescu}. To cut complexity, we must run symbolic
execution on models of the code, rather than the code itself. 
While it is natural to program the models also in C, as previous works do
\cite{symbex-dobrescu,symbex-vyas}, we show that C is
fundamentally ill suited for network symbolic execution, 
and the resulting models are too complex to analyze. 

We propose the Symbolic Execution Friendly Language (SEFL,
\S\ref{sec:sefl}), a network modeling language that we have developed
from scratch to enable fast symbolic execution. We have also developed
\symnet (\S \ref{sec:symnet-tool}), a tool that performs symbolic execution
on SEFL models of network boxes. Using \symnet and SEFL, we show it is
possible to run symbolic execution on large networks to understand network properties beyond simple packet reachability: safety of
tunnel configurations, MTU issues and  stateful processing (\S\ref{sec:evaluation_func}). 

The remaining challenge is to accurately model network functionality
in SEFL; this process requires expert input. We
have developed parsers that take switch MAC tables, router forwarding tables or 
CISCO firewall configurations (adaptive security appliance) and automatically
generate the corresponding SEFL models. Additionally, 
we have manually modeled a large subset of the elements from the Click
modular router suite \cite{click}: given a Click configuration
we can automatically generate the corresponding model. 
Finally, we have developed a testing tool 
that takes SEFL models and their runnable counterparts and automatically checks that
the model conforms to the actual implementation.

To evaluate \symnet, we have applied symbolic execution to
understand documented middlebox interactions\cite{op-middlebox}, to check our department's network
and the Stanford backbone. Results show
that \symnet  is more powerful than existing static analysis
tools, captures most real-life interactions in networks, with runtimes
in the order of seconds.
\section{Motivating examples}
\label{sec:examples}

Static analysis tools are enticing because they can help network
operators understand the operation of their deployed networks and they
can inform the correct deployment of updates. Static network analysis is
maturing: tools such as Header Space Analysis \cite{hsa} and Network
Optimized Datalog \cite{nod} have evolved from research and are now
being rolled into production. However, all existing tools have
limitations (see \S\ref{sec:related} for details), as they
choose different points in the tradeoff between the expressiveness of the
policy specification language, the network modeling language, the ease of modeling and the
checking speed. Furthermore, existing tools can't model widely used functions such 
as dynamic tunneling and encryption. 

In this section we discuss three examples of network functionalities we want to
statically analyze and highlight the difficulty in using existing
tools for this purpose.

\headline{Modeling tunnels.} Various forms of
tunneling are in wide\-spread use, sometimes deployed
by different parties. Can we statically verify that such tunnels
work correctly? We provide a simple example below, where E1 and E2 perform
IP-in-IP encapsulation and D1 and D2 decapsulation.

\begin{center}
$A \rightarrow E1 \rightarrow E2 \longrightarrow D2 \rightarrow D1 \rightarrow B $
\end{center}

Consider Header Space Analysis \cite{hsa} (HSA), the most mature
network static analysis tool today.  With HSA, the packet header is
modeled as a sequence of bits, where each bit can take values 0,1 or *
(don't care). Network functions are modeled as transformations of the
packet header. For instance, the IP-in-IP encapsulation could be
modeled as a 20 byte shifting of existing bits to the right and adding
the new IP header in the remaining space; the decapsulation will
perform the inverse operation.  Beyond reachability, we want to
statically answer the following basic question: are packet contents
invariant across this tunnel?  The answer is obviously yes, but HSA
cannot capture it: if the input header contains only * bits, the
output will also contain * bits, but this does not imply that
individual packet contents may not change. We can always feed a
specific packet to the model and check it is not modified, but to
ensure the invariant holds in general we must try all possible
packets---this won't scale. A symbolic packet is needed
instead---if the symbolic packet doesn't change, the property holds
regardless of its value.

We have also modeled the simple tunnel using the newer Network
Optimized Datalog tool \cite{nod}.  NOD can compute the
invariant, but modeling is cumbersome and limiting in many
ways. First, the models for D1 and D2 differ, despite the fact they
are running the same operation. D2 takes a packet with six header
fields (to ``remember'' the two layers of encapsulation): we cannot
reuse the D1 model for D2, nor the one from E1 for E2: we need to
create a new model instead.  In fact, network models in NOD depend not
only on the processing of the box, but also \emph{the network topology
  and the processing of other boxes}.  Additionally, models for boxes
operating at a lower stack level must also include higher level
protocol headers: for instance, a router will model not only its use
of the layer 3 fields, but also the upper layer protocols such TCP,
UDP or ICMP.  In summary, NOD modeling is extremely cumbersome when
network topologies are heterogeneous and run multiple protocols; this
is not an issue for datacenters where NOD's usage is targeted, but
will be a major issue in the wider Internet.

\headline{Classic symbolic execution} can be used to elegantly capture the
properties of tunnels, as we show in
\S\ref{sec:models}. Unfortunately, running existing symbolic execution
tools such as Klee \cite{klee} on the source code of network boxes
results in intractably long runtimes and memory usage. We now discuss
a series of examples that highlight these difficulties.

Dobrescu et al \cite{symbex-dobrescu} show that
symbolically executing an IP router implemented in C quickly becomes
intractable: when a packet with a symbolic destination address reaches the
router, the branching factor is at least as large as the number of
prefixes. Such branching is prohibitive when analyzing core routers that
have hundreds of thousands of prefixes in their forwarding tables.

To make symbolic execution tractable, we would like the branching
factor to depend on the number of links of the router instead; this is 
feasible, but \emph{we must write an optimized router \emph{model} for 
symbolic execution} (see \S\ref{sec:models}). 

\headline{Parsing TCP Options.}
Assume a network operator has deployed a stateful firewall and wishes to
know what options are allowed through in its current configuration; 
in particular, it wishes to understand whether a new IETF transport 
protocol might work in its network.

In Figure \ref{code:options} we show a C code snippet taken from a firewall
that processes the TCP options header field. The options field is
accessible via the \texttt{options} character array, and contains
\texttt{length} bytes. The middlebox we show allows a number of
widely used TCP options, and drops all other options by replacing them
with padding.

HSA or NOD can't model this example, but the operator could run \emph{klee} on the middlebox code
instead, providing a symbolic \texttt{options} field. 
By examining the options field after the firewall code, 
the operator can tell which options are allowed through and which
not. In Table \ref{tbl:klee_options} we present the number of resulting paths
and runtime of \emph{klee}, as we vary the length parameter whose max value is 40.
The results in Table \ref{tbl:klee_options} show just how costly symbolic execution on C code is, even on fairly
simple code snippets. 

\begin{table}
\centering
{\small
\noindent\begin{tabular}{|l|c|c|c|c|c|c|c|}
\hline
Length & 1 & 2 & 3 & 4 & 5 & 6 & 7\\
\hline
Number of paths & 3 & 8 & 19 & 45 & 106 & 248 & 510 \\
Runtime (s) & 0.2 & 3 & 20 & 109 & 501 & 2000 & 9500 \\
\hline
\end{tabular}
} 
\vspace{-0.1in}
\caption{Runtime of Klee on options parsing code.}
\label{tbl:klee_options}
\vspace{-0.2in}
\end{table}

\begin{figure}[t]
{\mateisize
\begin{verbatim}
unsigned char *ptr = &options[0];
unsigned char opcode,opsize;
while (length > 0) {
   opcode = *ptr;
   switch (opcode) {
     case TCPOPT_EOL: return True;
     case TCPOPT_NOP:        
         length--;ptr++;continue;
     default:
         opsize = *(ptr+1);
         if ((opsize < 2) || (opsize > length)){
             //nop everything!
             for (i=0;i<length;i++)
                ptr[i] = 1;
             length = 0;
             continue;
         }
         switch(_options[opcode]){
           case DROP: return False;
           case ALLOW: break;
           case STRIP:
              for (i=0;i<opsize;i++)
                  ptr[i] = 1;
        }
     }
     ptr+=opsize;length-=opsize;
}
\end{verbatim}
}
\vspace{-0.2in}
\caption{TCP Options processing code for a CISCO ASA box with default configuration.}
\vspace{-0.2in}
\label{code:options}
\end{figure}


\subsection{Towards a Solution}


When applying symbolic execution to C code, the number of branches in
the code exponentially increases the number of paths to be
explored. To make symbolic execution feasible, we need to drastically
simplify the code being symbolically executed. Unfolding loops and
executing both branches of an ``if'' instruction are techniques that
reduce the complexity of symbolic execution at the cost of increased
runtime \cite{overify}.  Such techniques, together with simpler data
structures allow verifying small pipelines of Click modular router
elements in tens of minutes, as shown by Dobrescu et
al. \cite{symbex-dobrescu}. 

Our target scale is two orders of magnitude higher: we aim to verify
networks containing hundreds or more elements in seconds. 
Ideally, the number of paths explored by symbolic execution should
be comparable to the number of paths in the network; in other words,
the model of any network box should not produce more paths than
the number of outgoing links from that box; for instance, in the
TCP options code in Figure \ref{code:options}, we
should have one or two 
paths at most.  This property would make
symbolic execution tractable even on very large networks. 
%

To achieve this target, we rely on the key observation that each \emph{path in network symbolic execution must
be tied to an active packet} passing through the network: if a codepath
does not result in packets, it should not be symbolically executed.
This implies that models of network boxes should only focus
on the paths that decide the fate of packets, leaving out any logging,
reporting, system checks, and so forth.

The C language does not have this property: a packet is just
one of many variables handled by the program, and dropping a packet
does not stop the execution of the box. Another fundamental problem is
the poor handling of data structures, as shown in our TCP options
example.

\section{Design Overview}

We take a radically different approach to enable large scale symbolic execution.
We need a new verification-driven modeling language that is imperative
---thus easier to program with---and that 
allows us to harness networking domain knowledge to simplify, as much as possible,
the task of the symbolic execution engine.
To this end, we have designed a novel language called SEFL 
that makes it possible to consider symbolic execution as part of
modeling rather than seeing it as a retrospective verification and
validation activity.

\begin{figure*}
{\footnotesize
\begin{tabular}{p{4.5cm}p{10cm}}
\emph{Instruction} & \emph{Description} \\
\end{tabular}

\begin{tabular}{|p{4.2cm}|p{12.8cm}|}
\hline 
\texttt{Allocate(v[,s,m])} & Allocates new stack for variable
v, of size s. 
If v is a string, the allocation is handled as metadata and the optional m parameter controls its visibility:
it can be global (default) or local to the current module. If v is an integer it is allocated
in the packet header at the given address; size is mandatory.\\

\texttt{Deallocate(v[,s])} & Destroys the topmost stack of variable v; if provided, the size s is checked against the allocated size of v. The
execution path fails when the sizes differ or there is no stack allocated for variable v. \\
\texttt{Assign(v,e)} & Symbolically evaluates expression e and assigns the result to variable v. All constraints applying to variable v in the
current execution path are cleared.\\
\hline
\texttt{CreateTag(t,e)} & Creates tag t and sets its value e, where e must evaluate to a concrete integer value. \\
\texttt{DestroyTag(t)} & Destroys tag t. \\
\hline
\texttt{Constrain(v,cond)} & Ensures that variable v always satisfies expression cond. The execution path fails if it doesn't. \\
\texttt{Fail(msg)} & Stops the current path and prints message msg to the console.\\
\hline
\texttt{If (cond,i1,i2)} & Two execution paths are created; the first one executes i1 as long as cond holds. the second path 
executes i2 as long as the negation of cond holds. \\
\texttt{For (v in regex,instr)} & Binds v to all map keys that match regex and executes instruction instr for each match. \\
\hline
\texttt{Forward(i)} & Forwards this packet to output port $i$. \\
\texttt{Fork(i1,i2,i3,...)} & Duplicates the packet and forwards a copy to each output port $i1,i2,..$. \\
\hline
\texttt{InstructionBlock(i,\ldots)} & Groups a number of instructions that are executed in order. \\
\texttt{NoOp} & Does nothing. \\
\hline
\end{tabular}
}
\vspace{-0.1in}
\caption{SEFL instruction set. }
\vspace{-0.2in}
\label{table:sefl}
\end{figure*}

A major design question regards the way packets are
modeled. With Header Space Analysis, headers have a fixed size, and
all possible layers have to be present at all time. This is fine for 
L2 boxes (such as Openflow switches), but won't work
in large, heterogeneous networks. Network Optimized Datalog models
packets as a collection of independent ``variables''; it can capture
tunnels to some extent, but it does not capture the physical layout of
packets, and the problems that encapsulation may bring
(e.g. interpreting the wrong part of the header, not knowing the
higher level protocol, etc.).

SEFL uses a packet layout that mimics real implementations.  As in
NOD, packet headers are variables, but each header has an absolute
offset at which it is allocated.  All SEFL headers must be allocated
individually, and allocation and deallocation commands include the
explicit size of the header field; no symbolic sizes are allowed to
ensure tractability.  Accesses to header fields must be
aligned, otherwise errors are thrown.
SEFL thus offers guaranteed \emph{memory safety} for packet headers and simplifies the 
symbolic execution engine which must not test for memory errors.

By design, in SEFL a \emph{packet is tied to an execution path},
and we use the terms path and packet interchangeably in this document. When
constraints applied to header fields are unsatisfiable, the execution
path fails altogether. 
This is in contrast with C where a packet header is just a regular variable 
and execution of a path terminates only on program exit. 
SEFL incorporates the following features:
\begin{itemize}
\item \textbf{Built-in map data structure.} \symnet offers  
a map datastructure that SEFL programs can use to create or retrieve values based on concrete string
keys. The map helps programmers avoid implementing complex data structures 
and their associated branching factor (such as those in Fig. \ref{code:options}). 

\item \textbf{Bounded loops}. The only loop instruction supported by SEFL is an iteration over s snapshot of the keys
in the native map data structure. This loop can therefore be unfolded and executed without any branching. 

\item \textbf {Dedicated path control instructions.} SEFL allows the user to explicitly drop a packet (path) based
on its contents without any branching. Multiple execution paths can be explicitly created with \texttt{fork}.
\end{itemize}

SEFL-coded models are by construction memory safe, have bounded memory usage and are guaranteed to terminate.

\section{SEFL Language}
\label{sec:sefl}



In table \ref{table:sefl} we list all the instructions
provided by SEFL, together with their parameters and
description. Every instruction implicitly takes as parameter the current
execution state (i.e. packet) and outputs a new execution state. The state includes
header variables and map entries (called metadata) together with their values and
constraints.

The \texttt{Allocate} and \texttt{Deallocate} instructions
create both header fields and metadata, depending on the 
parameter provided. If \texttt{v} is a string, the variable is metadata and is
not aligned in any way, and memory safety checks do not apply. \texttt{v} acts as a key
in the map managed by SymNet.
If \texttt{v} is an integer (or an expression that evaluates to an integer), it is treated like a header
with the associated memory checks. 

To simplify access to header fields and to enable layering, 
any number of \emph{tags} can be defined. The programmer can use indexed addressing
based on the tags and a fixed offset rather than absolute addresses to access a header field.

SEFL has instructions to manipulate tags. New tags can be created 
at absolute values (used when the packet is created), or relative to other tags (used
for encapsulation of an existing packet). 
Tags can be defined dynamically,
pointing to addresses where layer two, three and four headers start. 
The code below performs IP header encapsulation:

{
\vspace{-0.05in}
\small
\begin{verbatim}
    CreateTag("L3",Tag("L4")-160)
    Allocate(Tag("L3")+96,32)   //IP src
    Assign(Tag("L3")+96,
             ipToNumber("192.168.1.1"))
    Allocate(IpDst,32)          //IP dst
    Assign(IpDst,ipToNumber("8.8.8.8"))
\end{verbatim}
\vspace{-0.05in}
}

The notation to access the IP source address field is rather wordy. To 
make programming easier, we have defined shorthands for all header fields
that we work with: {\small \verb!Tag("L3")+96!} becomes \texttt{IpSrc}. 
The code to initialize the destination address field uses this shorthand, 
and is also easier to read. The decapsulation code does the opposite:
first, header fields are deallocated then the L3 tag is destroyed.

SEFL includes two instructions that constrain the execution of the
current path, that have no direct correspondent in C. \texttt{Fail}
stops the current execution path and prints an error message.
\texttt{Constrain} applies a constraint to a variable, stopping the
current path if the constraint does not hold. 
\texttt{Constrain} allows programmers to model filtering behaviour without branching. 
Below we show the SEFL and C code to drop non-HTTP packets.

{\small
\vspace{-0.05in}
\begin{verbatim}
Constrain(TcpDst==80)       if (p->dst_port==80)
                               free(p);
\end{verbatim}
\vspace{-0.05in}
}

\noindent The C code results in two execution paths if the dst\_port 
field is symbolic, while SEFL only adds the \texttt{TcpDsp==80} constraint
to the current path.

\texttt{If} forks the current execution state. On one path, it applies the constraint and executes \texttt{instr1}.
On the ``else'' branch is applies the negated constraint and executes \texttt{instr2}. If more than one
instruction must be executed on any branch, an \texttt{InstructionBlock} should be used that
groups more instructions into a single compound instruction. If any branch is empty, \texttt{NoOp} 
can be used instead.

\texttt{For} iterates over the keys in the map that match a given pattern. The code does not branch: a snapshot of the keys
is taken and the loop is unfolded before execution.

\section{Symbolic execution}
\label{sec:symnet-tool}

\begin{figure*}[t]
\begin{minipage}{0.2\textwidth}
\includegraphics[width=1.05\textwidth,trim=2.1cm 9.5cm 15cm 1cm,clip=true]{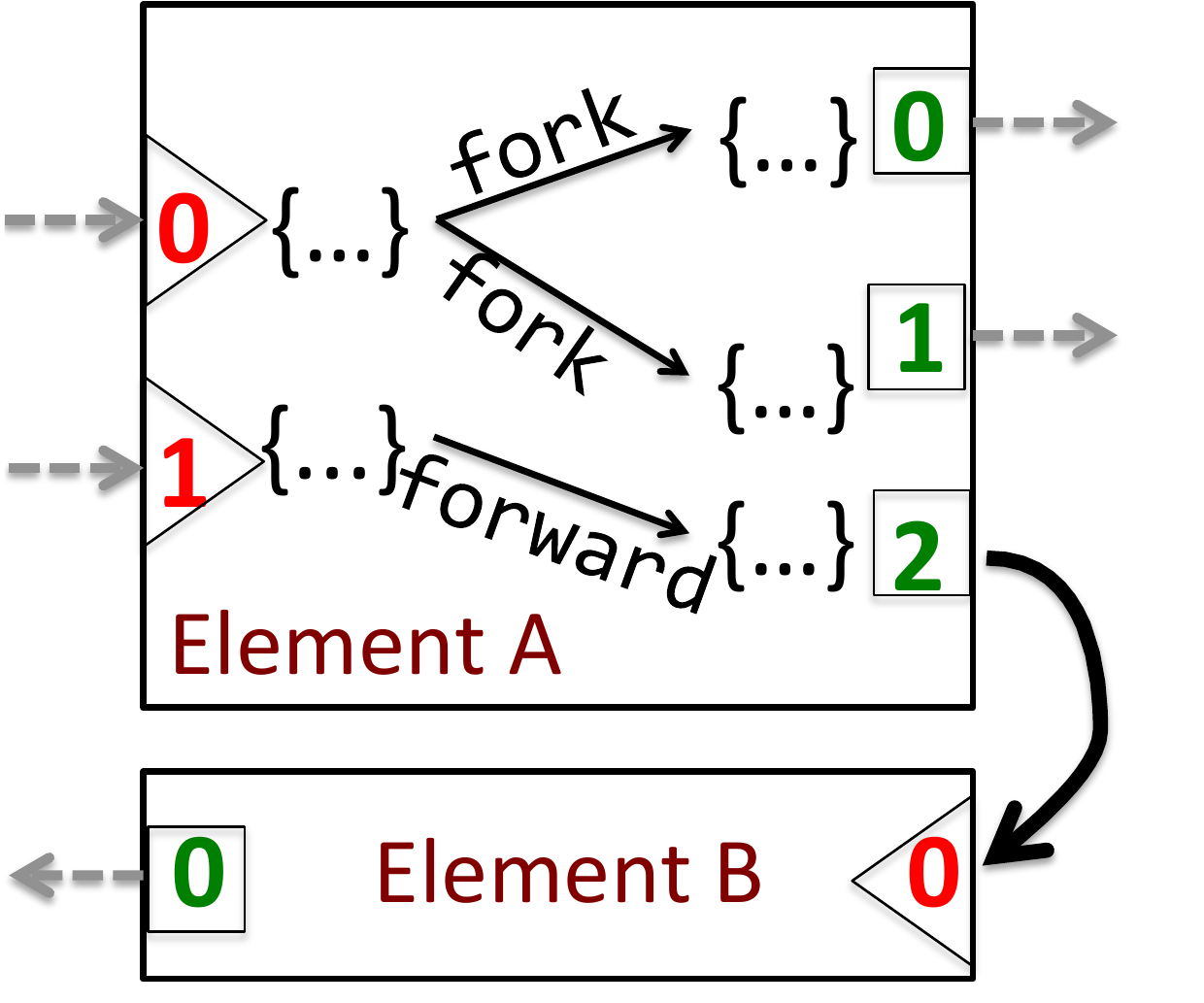}
\caption{Network model with two elements.}
\label{fig:elem}
\end{minipage}
\begin{minipage}{0.56\textwidth}
\includegraphics[width=\textwidth,trim=3.0cm 11cm 2cm 1cm,clip=true]{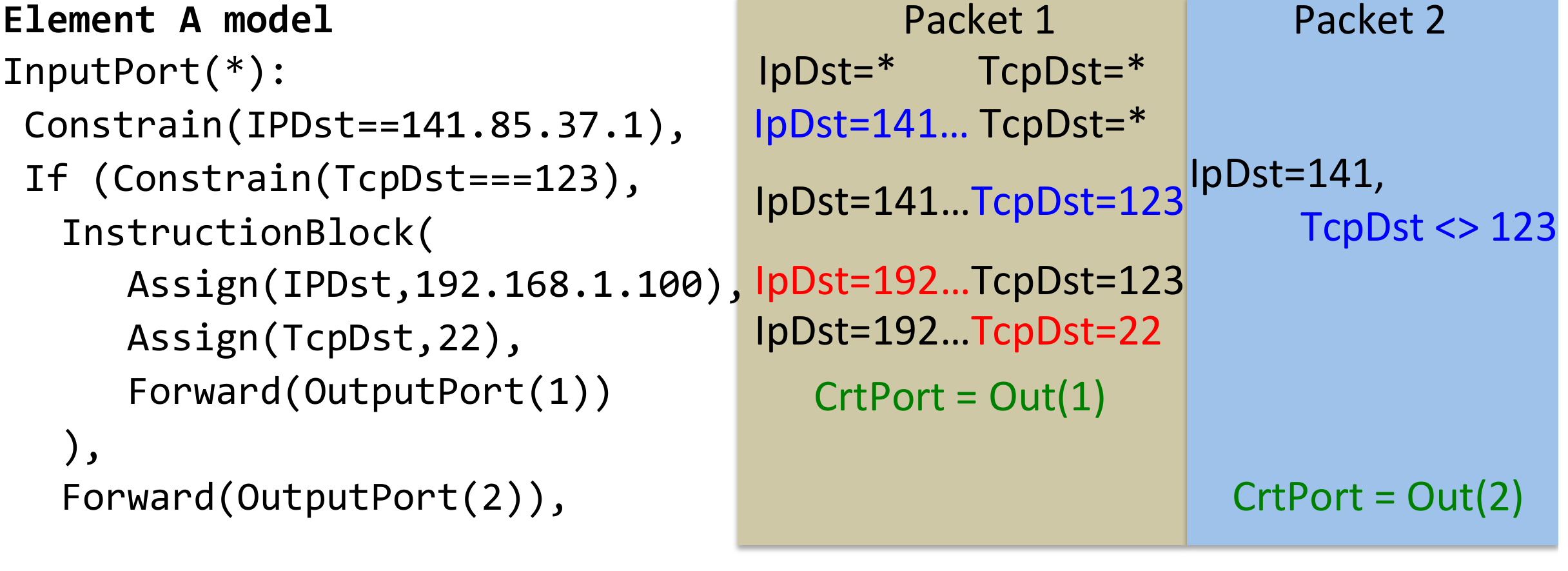} 
\vspace{-0.3in}
\caption{Symbolic execution with \symnet. The tool keeps a per-path value stack
and assignment history for each variable.}
\label{fig:symnet-example}
\end{minipage}
\begin{minipage}{0.21\textwidth}
\includegraphics[width=\textwidth,trim=1.5cm 8cm 15.5cm 1cm,clip=true]{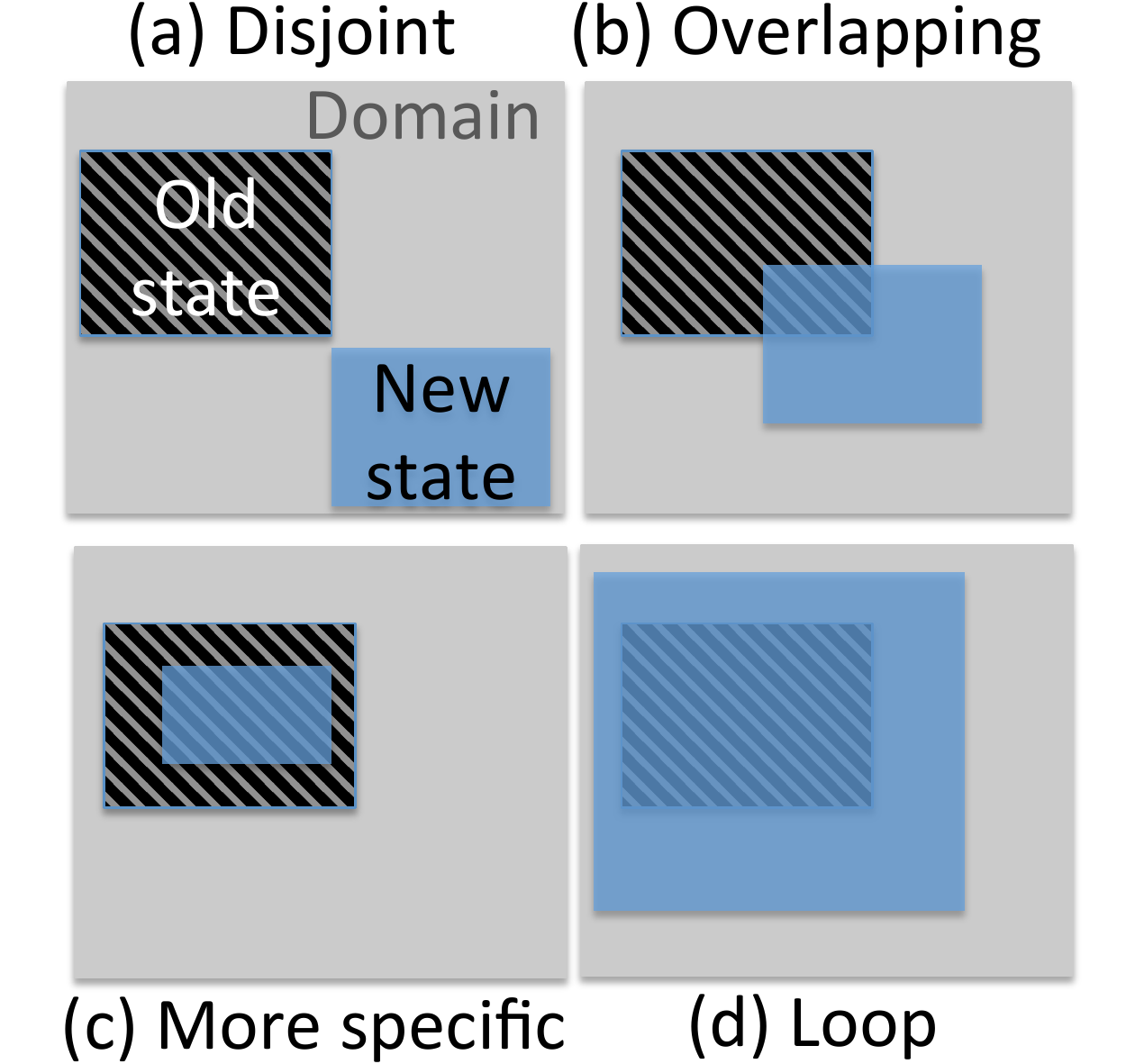}
\vspace{-0.25in}
\caption{Loop detection algorithm example.}
\label{fig:loop}
\end{minipage}
\vspace{-0.25in}
\end{figure*}

Our symbolic execution tool is called \symnet and consists of 
15KLOC of Scala code. 
To analize a network configuration, \symnet requires as input the descriptions
of all the network elements and their connections. 
Each network element has input and output ports, as shown in 
Fig. \ref{fig:elem}: input ports are shown with a triangle and output
ports are shown as rectangles. Connections are unidirectional from output to input ports,
so we need two pairs of ports and two links for bidirectional connectivity. 
Providing a model for a network element means specifying the number
of inputs and output ports and associating a set of SEFL instructions
to each port. 

\symnet starts execution by creating an initial empty packet, with no header fields or metadata,
and then executes code to create a symbolic packet of the given type (e.g. TCP). The packet is injected 
into an input port specified by the user, e.g. port 0 of element A in Fig.\ref{fig:elem},
where \symnet executes the associated instructions.

\symnet  instructions take as
input an execution path and can modify its associated state, spawn new
execution paths or both. The instructions on port 0  may forward the packet to one of the output ports by using the 
\texttt{forward} or \texttt{fork} instructions. After the output port instructions are executed,
if the state is feasible and there is an outgoing link from the output port, the packet is processed
at the next input port. 

Values in \symnet can be concrete or symbolic; each value has a unique
identifier. For each value, on each path, \symnet holds a list of
constraints that apply to that value. Assignment operations modify the
top of the current value stack.  A path finishes execution when the
\texttt{Fail} instruction is called, when a constraint does not hold
on any of its variables, or when it reaches a port with no outgoing
links.

\symnet enhances general purpose symbolic execution in two major ways.
First, the state contains the map that stores variable names (or
memory addresses) as keys and their associated \emph{value stacks},
instead of simple values.  Allocation and deallocation instructions
push and pop a whole value stack.  This allows programmers to quickly
``mask'' the current value of a variable and restore it later with
ease. Secondly, \symnet keeps a complete history of the values
associated with a symbol which allows it to detect network-wide loops
and to check for field invariance across network hops.

\symnet  is also considerably simpler than tools like Klee: it does not use heuristics to prioritize paths to explore
  because our target is finding all the possible execution paths through the
  network, not just covering all instructions of the network model with at least
  one execution path. 
  Additionally, \symnet  (via SEFL) only supports simple expressions (referencing,
  subtraction, addition, negation), and this greatly reduces state representation
  complexity.

To better understand \symnet execution, look at the toy example in
Figure \ref{fig:symnet-example} showing an implementation of port
forwarding in element A.  The code is executed when packet 1 reaches any of the input
ports of A. The packet enters with
header fields \texttt{IpDst} and \texttt{TcpDst} set to symbolic values.
The \texttt{constrain} instruction forces packets to be destined to a specific IP address.  The
\texttt{If} instruction then creates another packet, packet 2.
All the state of packet 1 is replicated to packet 2 (in fact, it is
shared with a copy-on-write mechanism).  Next, packet 1 gets the
\texttt{TcpDst==123} constraint added and checked by a constraint
solver (Z3 \cite{z3}, in our case). The constraint is satisfiable, so
packet 1 is propagated further to the assignment instructions which
rewrite both the destination address and ports of the packet before
forwarding it to output port 0. Packet 1 is now done and \symnet will
return to Path 2, first adding the negated constraint
\texttt{TcpDst!=123} and then asking the solver if it is satisfiable.
The packet goes to output port 2 and to element B (port 0).
\section{Network Verification}
\label{sec:verify}

\headline{Reachability.} It is straightforward to check reachability
in a network modeled with SEFL. A symbolic packet is 
injected at the desired source port, and this packet is then
propagated through the network by \symnet. At each port reached by the
symbolic packet, we can inspect the values of and constraints on the header
variables to discover which packets are allowed, 
what input packets can reach the output, and how the packets
look like at the output, on \emph{all the execution paths that reach that port}.

In the example in Figure \ref{fig:symnet-example}, a single path reaches output port 1.
By examining the history of this path, we can conclude the port is 
reachable only when the incoming packet satisfies  \texttt{TcpDst==123}, and
the outgoing packet will have its destination address and port overwritten.
Output port 2 is also reachable by packets that satisfy \texttt{TcpDst!=123}.

\headline{Loop detection.} The loop detection algorithm relies on the
reachability algorithm and we run it at every port in the network.
When a new port is visited, we save the current execution state (all variables 
and all constraints). When the same port is revisited, the current state of each variable 
is compared to \emph{all its previous states}.  

Figure \ref{fig:loop} shows how the current state compares to the old ones. A loop exists
only when the new state contains all possible values in the old state (case (d) in the figure).
Say the set of all constraints for the old state is $o$, and for the new state is $n$.
To check for loops, we invoke the solver with the constraint \texttt{!n \& o}. In Fig.\ref{fig:loop}, this constraint asks 
the solver to find a point included in the old state that is not contained in the new state. If the solver
returns an example, there is no loop, otherwise a loop exists. In our figure, the solver will find counter-examples
for cases (a)-(c), but not for (d).

The loop detection algorithm is generic and can capture different kinds of loops. If
we apply it to the entire state, the algorithm will not capture traditional forwarding loops
because the TTL field will always decrease and thus the state will be
different.  To capture such loops, we must apply the same algorithm but
only consider destination and source IP addresses when comparing states.

SEFL models (programs) are inherently bounded in space - there is no
recursion or heap allocation. We have
also proven that all SEFL programs have bounded execution. Loops can only appear as a result of the network
topology, but are captured by our loop detection algorithm.

\headline{Invariants.} By checking the value stack of the destination
address field on output port 2, we find that it is bound to the same symbolic
value that was set when the path entered the input port. In Fig. \ref{fig:symnet-example} 
the \texttt{TcpDst} and \texttt{IpDst} fields are invariant on path 2
as long as \texttt{TcpDst!=123}.

\headline{Header visibility.} By analyzing the value stack of a header
field at an intermediate point, we can understand whether the value
read is the same as that set by the source or seen by the
destination. Visibility tests allow us to check whether firewalls 
and endhosts see the same headers. 

\headline{Header memory safety.} When creating or destroying header fields,
accesses are indexed through tags. If the tags are set incorrectly, of if the
program wrongly assumes the location of headers, the execution path will
fail. This has allowed us to catch various encapsulation 
problems in buggy models.






\section{Modeling networks with SEFL}
\label{sec:models}

\begin{figure}
\centering
\includegraphics[width=.8\columnwidth, trim=1cm 10cm 1cm 1cm,clip=true]{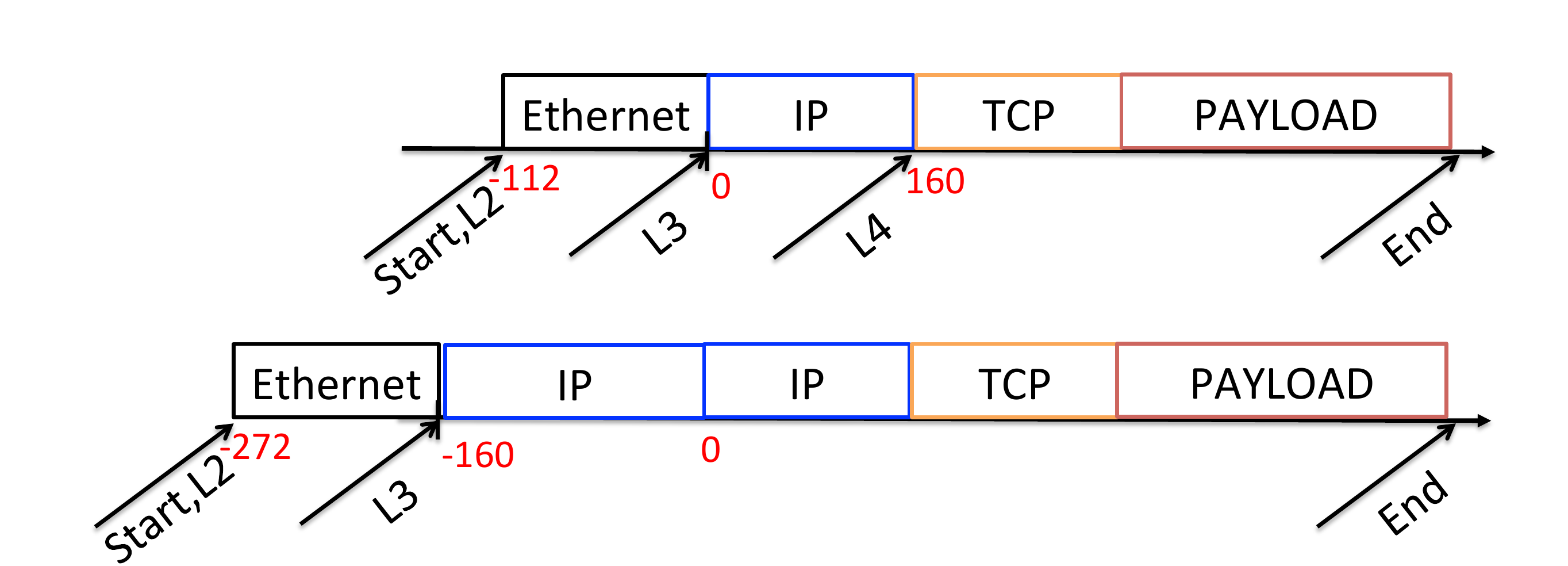}
\vspace{-0.15in}
\caption{\symnet  packet modeling uses the same physical layout as real packets.}
\vspace{-0.2in}
\label{fig:header}
\end{figure}

We begin with more detail on how we model packets in Figure \ref{fig:header}. At
the top of the Figure a TCP packet is encapsulated with IP and
Ethernet headers.  Packets always have the Start and End tags set; all
the other tags are set as packets move through the (modeled)
stack. Layer tags are always allocated relative to other tags; the
start and end tags start at 0 by convention when a symbolic packet is
created.
The bottom packet in Figure \ref{fig:header} is an IP-in-IP encapsulated packet. The L4 tag is
not set in this case; this will be set only in the IP decapsulation code. Any accesses to L4 fields before
the L4 tag will fail, stopping the associated path. 
SEFL network models support protocol layering natively, shielding lower layer models from semantics higher up the stack. 

The code below models a packet received from an Ethernet network interface; it first
sets the L2 tag and only allows IP packets destined for a certain MAC address.
Once the L2 tag is set, the L3 tag can be set by adding 112 bits to it:

{\small
\vspace{-0.1in}
\begin{verbatim}
    InputPort(0):
      CreateTag("L2",Tag("Start"))
      Constrain(EtherDst,==00:aa:00:aa:00:aa)
      Constrain(EtherProto,==0x0800)
      CreateTag("L3",Tag("L2")+112)
      Forward(OutputPort(1)) 
\end{verbatim}
\vspace{-0.1in}
}


\headline{Modeling switch behaviour.} To model switches, we
have written a parser that takes a snapshot of CISCO switch MAC
tables (containing MAC, VLAN and output port tuples) and creates a SEFL model. 
To reduce branching, our
model simply groups all MAC addresses that should be forwarded on the
same output port in a single constraint. A possible switch model is the one
below which we call ``ingress'' because the filtering code is applied on the
input ports:

{\small
\vspace{-0.05in}
\begin{verbatim}
    InputPort(*):
      If (Constrain(EtherDst==MAC11 |
              EtherDst==MAC12 | ... ),
         Forward(OutputPort(1)),
         If (Constrain(EtherDst==MAC21 | 
                EtherDst==MAC22 | ... ),
               Forward(OutputPort(2)),
               Fail("Mac unknown")))
\end{verbatim}
\vspace{-0.05in}
}

When we run the code above with a symbolic \texttt{EtherDst}, it will
result in as many execution paths as the number of output ports of the
switch, which is optimal. However, it will generate more constraints
than needed: a path taking an \texttt{else} branch will include the
negated constraints for the \texttt{if} branch together with the
constraints for the current output port.  To avoid these additional
constraints, we can write the switch using an \emph{egress} filtering
model:

{\small
\vspace{-0.1in}
\begin{verbatim}
    InputPort(*):
        Fork(OutputPort(1),OutputPort(2),...)
    OutputPort(1):
        Constrain(EtherDst==MAC11 | MAC12 ...)
    OutputPort(2):
        Constrain(EtherDst==MAC21 | MAC22 ...)
\end{verbatim}
\vspace{-0.1in}
}

The egress model has both optimal branching and a minimum number of constraints.
The egress models is correct as long as the constraints are mutually exclusive, which always holds for MAC tables
due to the spanning tree algorithm. 

\headline{Modeling an IP Router.} At first sight, it seems we should be able to use the same approach
to model an IP router, but this is not true in the following forwarding table:

{\small
\vspace{-0.1in}
\begin{center}
\begin{tabular}{|ccc|}
\hline
Prefix & & Output Interface \\
\hline 
192.168.0.1/32 &$\rightarrow$& \texttt{If0} \\
10.0.0.0/8     &$\rightarrow$& \texttt{If0} \\
\hline
192.168.0.0/24 &$\rightarrow$& \texttt{If1} \\
10.10.0.1/32   &$\rightarrow$& \texttt{If1} \\
\hline
\end{tabular}
\end{center}
\vspace{-0.1in}
}

If we simply group the rules per output interface and apply them
using \texttt{If} instructions in the order above, the resulting
forwarding will not use longest prefix match for destination address
10.10.0.1, which will be forwarded wrongly on \texttt{If0}.
The most obvious solution is to have one \texttt{If} instruction for each
prefix and ensure that for all overlapping prefixes, more specific matches are checked first. However,
this creates as many branches as the number of prefixes in the routing table. In our example we would have 
four branches, but for core routers this means hundreds of thousands of branches.

A better algorithm is the following. If prefix \emph{a} is more specific than
prefix \emph{b}, create the following constraint for \texttt{b}: \texttt{!a \& b}. This ensures
that the more specific prefix \texttt{a} does not match. We can now group all rules that have the same
output interface as in the switch case; the number of resulting paths drops from the number of prefixes
to the number of links of the router, which is again optimal. We also rely on egress filtering 
to reduce the number of constraints.

\headline{Modeling a Network Address Translator.} NATs are ubi\-quiously
deployed as operators come to grips with the IPv4 address space
shortage. NATs modify the source IP address and source port for
outgoing packets and apply the reverse mapping for incoming packets.
NATs are harder to model: they keep per flow state to ensure
incoming traffic is only allowed if it is related to outgoing traffic
the NAT has seen. In addition, the list of available ports at a NAT is
a global variable, and the port assigned to a new connection will depend
on many external factors, such as the number of active connections, the random
number generator, and so forth.

To model the NAT we observe that the exact port number assigned by 
the NAT is quasi-random, and network operators treat it as such. Therefore
it makes no sense to model the algorithm used to choose a port for a new
connection; this would simply not scale. Instead, the newly mapped port 
will be a symbolic variable with allowed values in the NAT's port range.
To make the NAT ``remember'' a mapping for the flow, we save the NAT state
using metadata: 

{\small
\vspace{-0.05in}
\begin{verbatim}
    InputPort(0):
       Constrain(IPProto,==6) //only do TCP
       Allocate("orig-ip",32,local)
       Allocate("orig-port",16,local)
       Allocate("new-ip",32,local)
       Allocate("new-port",16,local)
       Assign("orig-ip",IpSrc) //save initial addr
       Assign("orig-port",TCPSrc) //save initial port
       Assign(IpSrc,"...")        //perform mapping
       Assign(TcpSrc, SymbolicValue())
       Assign("new-ip",IpSrc) //save assigned addr
       Assign("new-port",TcpSrc) //save assigned port
       Forward(OutputPort(0))
\end{verbatim}
\vspace{-0.05in}
}

On the return path, the code restores the original mappings only if the metadata
is present and matches the mapping the NAT has assigned to this flow:

{\small
\vspace{-0.05in}
\begin{verbatim}
    InputPort(1):
       Constrain(IPProto,==6)
       Constrain(IpDst,=="new-ip")
       Constrain(TcpDst,=="new-port")
       Assign(IpDst,"orig-ip")
       Assign(TcpDst, "orig-port")
       Forward(OutputPort(1))
\end{verbatim}
\vspace{-0.05in}
}

The NAT uses local metadata to ensure that multiple instances of the code can be run cascaded. Local 
metadata will ensure each NAT instance stores and retrieves its own values. 
Our NAT does not create any branches - the return packet is allowed if it contains the mapping,
or dropped otherwise. 
The NAT code is a faithful model of the real thing. 

The technique we used to model the NAT---storing per flow state inside the packet---we also 
used to model other similar boxes including stateful firewalls and firewalls that randomize the initial 
sequence number of TCP connections. The same technique can be applied wherever the per-flow state 
is independent across flows. Under this (admittedly strong) assumption, symbolic execution can verify
large networks with stateful middleboxes without state explosion. 

\headline{Modeling TCP options parsing.} Our models for routers and
switches are exact---there is no simplification compared to the real
code as our model accurately mimics the behaviour of the code. To make 
the options parsing code symbolic-execution friendly
though, we need to simplify it. The main problem with that code is
that it includes a for loop with an unknown number of
iterations, and the code has branches in the loop body.
Simplifying the code means we cannot capture all the properties of the
original code. 

To understand the difference in properties that can be
proven on the original C code and our SEFL model, we have performed a thorough
analysis in \S \ref{sec:eval:parse}. We find that traditional symbolic 
execution can prove memory safety and bounded execution on the real code,
albeit only on a small subset of inputs. When answering higher level
questions such as ``which options are allowed through?'', the answers
can be wrong because only a small part of the options field can be checked
in reasonable time.




We create a model of the options parsing code that is amenable to fast symbolic execution.
It is based on the observation that the order in which options are placed in the options field does not matter
---the code allows known options and strips everything else. This suggests that we
can modify the list data structure with SymNet's in-built map.
More concretely, each possible TCP option \emph{x} (where x is in
between 2 and 255) will have a corresponding metadata variable called
``OPTx'' that can take values 1 or 0, modeling whether that TCP
option is enabled or not.
The option length and
body will be held in metadata variables ``SIZEx'' and ``VALx''
respectively. In a sense, our model pre-parses the byte representation
of the options and stores it in the packet metadata, allowing quick access to the options.

A snippet of the firewall options code is given in Figure \ref{options:sefl}. Stripping options
is simply a matter of setting the associated metadata to 0, regardless of the initial value---hence the
is no branching involved. The SACK\_OK is stripped only for HTTP traffic.
The code then always sets the MSS option, and rewrites its value to be at most 1380. 

\begin{figure}
{\small
\begin{verbatim}
Assign("OPT30",ConstantValue(0)),
If (Constrain(TcpDst==80),Assign("OPT4",0),NoOp),
Assign("OPT2",1),
Assign("SIZE2",4),
If (Constrain("VAL2">1380),Assign("VAL2",1380),NoOp)
\end{verbatim}
}
\vspace{-0.15in}\caption{ASA options parsing code modeled in SEFL. }
\vspace{-0.2in}\label{options:sefl}
\end{figure}

Our model can accurately tell which options are
allowed through and under what circumstances. The code has few branches and is thus
cheap to symbolically execute, and captures more properties than 
Klee on the C code (see \S \ref{sec:eval:parse}).


\headline{Modeling Encryption.} Encrypted tunnels are being deployed more and more. 
We need to capture two properties: first, no network box can read the original contents of the payload once it is encrypted.
Secondly, if we decrypt using the same key that was used for encryption, we will retrieve the original payload.
As in the NAT case, predicting the way the ciphertext will look is not important for our model.
All that matters is that the original content is not available after encryption. We could
use the following code snippet to encrypt with key $K$, where K is a parameter.

{\small
\vspace{-0.05in}
\begin{verbatim}
     InputPort(0):
       Allocate("Key")
       Assign("Key",K)
       Allocate(TcpPayload)
       Assign(TcpPayload,SymbolicVariable)
       Forward(OutputPort(0))
\end{verbatim}
\vspace{-0.05in}
}

The decryption only proceeds if the key matches:
{\small
\vspace{-0.05in}
\begin{verbatim}
     InputPort(0):
       Constrain("Key",==K)
       Deallocate(TcpPayload)
       Forward(OutputPort(0))
\end{verbatim}
\vspace{-0.05in}
}

Despite its simplicity, the code above has the two properties we seek. Any box reading the TCP payload
after encryption will only see a novel unbounded symbolic variable, not the original contents. Only 
using the proper decryption key will retrieve the original contents.

\subsection{Ready-made network models}
\vspace{-0.05in}
Writing models in SEFL requires expert input. It cannot be reasonably assumed that network
administrators have the time or the expertise needed to perform this task, yet they are the
main beneficiaries of the tool. To make \symnet easily usable, we have created
parsers that take configuration parameters and/or runtime information from well known network elements 
and output corresponding SEFL models. All the user has to do is place all these files in a single 
directory, together with a file describing the links between the boxes. Then, the user can run \symnet
by specifying an input port to start the reachability and loop detection analysis. 

The output of the tool is the list of explored paths in json format. For every path \symnet lists all
variables and their constraints at the end of the execution as well as all the instructions and ports
this path has visited. The user can check these paths either manually or with standard tools (e.g. grep)
to see if certain undesired paths have been visited, etc. 

\headline{Switches and routers.} \symnet generates models from snapshots of switch MAC or router forwarding tables.

\headline{Click modular router configurations.} We have modeled in
SEFL a large subset of the elements of the Click modular router. This
exercise has served two main purposes: first, it allowed us to
understand whether SEFL's limited instruction set is sufficient to
model a wide range of functionality. Second, we use the Click elements to build more complex boxes
such as firewalls, NATs and even a CISCO application
security appliance. Our parser takes the Click configuration file,
generates a model for each individual element and then connects these models according to the
config file. 

\headline{Openstack Neutron configurations.} Openstack Neutron allows cloud tenants to specify
at an abstract level the networking configuration of their VMs before they are instantiated. Users can specify
firewalls, router configurations, virtual links, etc. We have written an Openstack plugin that takes the
router and firewall configurations and translates them into SEFL models. These could be used to check 
reachability before deployment, or to check that the actual deployment matches the user's intent; we are still
working on integrating the results from symbolic execution back into Neutron to make it easily available to 
the users.
\subsection{Modeling a CISCO ASA}
\label{sec:asa}
\vspace{-0.05in}To analyze our department's network, we must model its core device: a Cisco ASA
(Adaptive Security Appliance) 5510 firewall, henceforth called ASA.
It combines basic layer-2 capabilities such as switching and VLAN
segmentation, with static \& dynamic NAT and stateful packet
inspection and filtering. 
How the latter is achieved, and the internal cuisine of the inspection
process is the major challenge in accurately modeling ASA
behavior. For instance, by default, the ASA will intercept TCP
connections, and act as server until the connection is actually
established. This feature protects machines behind the ASA from TCP
SYN floods. However, documentation is very sparse and generic, making
it difficult to understand the exact behaviour. 


To understand how the ASA processes traffic, we developed a
``black-box" testing environment, using the same ASA model. 
We connected the real ASA to two machines, generically termed
\emph{outside} and \emph{inside}. Using Click, we ran test sequences
consisting in TCP, UDP, ICMP and simple (raw) IP packets, through the
ASA and observed the outputs. 

The next step is to generate the model. We could generate SEFL
directly, however SEFL is not executable in practice. We chose to
generate a Click configuration that models the ASA instead. \symnet  can execute
this configuration because it has models for each element; the bonus
of Click modeling is that we can potentially run the ASA in software,
by simply instantiating it.
After weeks of black box testing, we implemented a tool that parses the ASA configuration
file and generates a Click ASA model automatically.
We experimented with different ASA configurations produced by our
tool, running the same test sequences through the real ASA, as well as
the Click model. The default ASA configuration includes static and dynamic
NAT, traffic filtering and basic TCP protocol inspection. Our model captures 
layer 2 and 3 behaviour, NATs.
The Click configuration uses standard elements, with the exception of a new element called TCPOptions which 
implements the options filtering code.

The resulting (simplified) Click packet pipeline can be summarized as
follows: (i) \emph{ingress static nat}: if the packet matches an
incoming static NAT rule, then it is modified accordingly; no state is
preserved for such packets, (ii) \emph{TCP inspection}: if the packet
is the response of an active TCP connection, it is forwarded to
destination directly, and appropriate dynamic NAT rules applied, if
this is the case; otherwise (iii) \emph{filtering}: appropriate
filtering rules are applied; (iv) TCP connections are stored, and
dynamic NAT mappings are inserted, for TCP packets; finally (v)
\emph{egress static nat}: if the packet matches an outgoing static NAT
rule, then it is applied. Finally, all TCP packets are parsed by 
the TCPOptions element with code very similar to that in Fig.\ref{code:options}.

\section{Evaluation}
\label{sec:evaluation}

Our evaluation has two major parts. First, we want to understand
whether our approach tailoring network models for symbolic execution
pays dividents in terms of runtime and accuracy. 
Secondly, we wish to understand whether symbolic
execution can help us find new types of bugs not captured by existing
tools such as HSA or NOD. We explore two network deployments: our department's network
containing a Cisco middlebox, one router and 15 switches and an 
enterprise middlebox deployment \cite{op-middlebox}.

\subsection{Performance evaluation}

We ran experiments using our \symnet prototype on a quad-core Intel i5
machine with 8GB of RAM, testing models of network functionalities
such as routers, switches, firewalls, options parsing and Click elements.

\headline{Symbolic execution of a switch model.} We used a snapshot of the MAC
table of the core switch in our department network (see Fig \ref{fig:csnet}) 
with four hundred entries to model its behaviour. We built three models of
the switch:
\begin{itemize}
\item \textbf{Basic}: a lookup table, with one entry per MAC, applied on ingress. This is the same as running a generic
symbolic execution tool on switch forwarding code. 
\item \textbf{Ingress}: group MACs going to the same output port, apply filtering and take switch decision on input. 
\item \textbf{Egress}: fork traffic to all outgoing ports and apply per-port restrictions on egress. 
\end{itemize}

\begin{figure*}[t]
\begin{minipage}{.33\textwidth}
\includegraphics[width=1.1\textwidth]{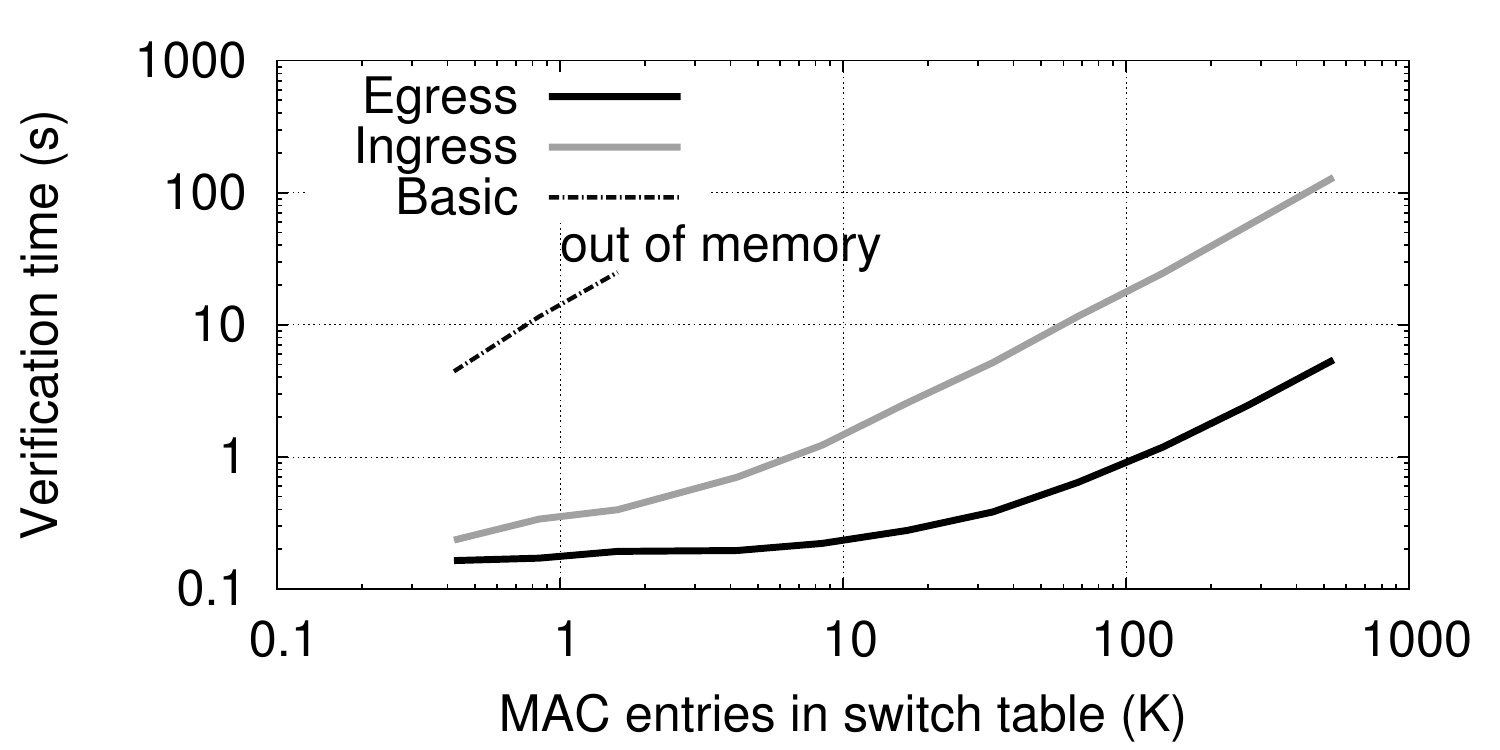}
\vspace{-0.25in} 
\caption{Symbolic execution of different switch models.}
\label{fig:switch}
\end{minipage}
\begin{minipage}{.33\textwidth}
\centering
\small
\begin{tabular}{|c|c|c|c|}
\hline
\textbf{Prefixes} & \textbf{Basic} & \textbf{Ingress} & \textbf{Egress} \\
\hline
1600 & 25s & 2.1s & 0.4s \\
\hline
62500 & DNF & 23.1s & 5.6s \\
\hline
188500 & DNF & DNF & 18s \\
\hline
\end{tabular}
\vspace{-0.1in}
\captionof{table}{Core router analysis.}
\vspace{+0.1in}
\label{tbl:router}
\begin{tabular}{|p{2cm}|c|c|c|}
\hline
 &  HSA  & \symnet  \\
\hline
Generation Time &  3.2min &  8.1min \\
Runtime &  24s &  37s \\
\hline
\end{tabular}
\vspace{-0.1in}
\captionof{table}{Comparison to HSA}
\label{tbl:hsa}
\end{minipage}
\begin{minipage}{.33\textwidth}
\centering
\mateisize
\begin{tabular}{|c|cc|cc|}
\hline
\textbf{Property} & \textbf{Klee} & \textbf{\symnet} \\
\hline
\emph{Runtime} & 1h & 1s \\
\hline
\emph{Bounded execution} & yes upto 6B & no \\
\hline
\emph{Memory safety} & yes upto 6B & no \\
\hline
\emph{Invalid Length} & yes upto 6B & less precise \\
\hline
\emph{SackOK,MSS,WScale} & yes upto 6B & yes \\
\hline
\emph{Timestamp,Multipath} & incorrect & yes \\
\hline
\emph{Combinations of options} & incorrect & yes \\
\hline
\end{tabular}
\captionof{table}{Comparison between Klee and \symnet on TCP options firewall code.}
\label{tbl:options}
\end{minipage}

\vspace{-0.2in}
\end{figure*}

We inject a packet with a
symbolic destination MAC address and execute the switch code and
forward all the resulting paths on the corresponding output ports.
We measure wall-clock execution time and the number of paths,
time spent in and number of calls to the constraint solver. 

Figure \ref{fig:switch} plots the \symnet runtime as we increase the
number of MACs from 440 to 500,000.To generate more entries in
the MAC table, we duplicate existing entries as many times as
needed; each entry gets a unique destination MAC address. In
all tests, more than 90\% of time is spent in Z3.

The basic model generates as many paths as entries in the lookup table. 
Each path will have its own instance of the solver and a set of constraints, and this
has a large memory overhead: it takes 10 seconds to execute a 1000-entry MAC 
table, and close to 8GB of RAM. Beyond this size, the RAM on our machine is insufficient.

The ingress model groups MAC addresses by output port, and results in
as many paths as output ports in use on our switch (20). However, the
path constraints are quite complex: the first path will contain the
allowed MACs, the second path will contain its own allowed MACs and
the negated constraints on the first path, and so forth. In fact, the
total number of constraints grows quadratically with the number of
switch ports: it takes 2 minutes to symbolically execute a switch code
with 480,000 entries.

The egress model also groups MACs per output port, but avoids the
negation by forking the initial flow and applying the constraints
independently on each port. In this case the total number of
constraints is the same as the number of entries, and this gives a big
benefit in runtime: it takes only 5s to execute 480 thousand
entries. This is the switch model we use in the remainder of our
evaluation.

\headline{Router.} 
We used a publicly available snapshot of the forwarding table of a
core router containing 188.500 entries \cite{stanford-data}. We generated the model by checking for
overlapping prefixes and adding constraints to ensure the per-port
constraints are mutually exclusive: this results in 183000 additional
constraints to a total of 371.000 prefix checks (or 722.000
inequalities). Model generation takes 8 minutes.

We run tests with 1\%, 33\% and 100\% of the prefixes by injecting an
IP packet with a symbolic destination IP address into the three
models: basic, ingress and egress. We provide the results in Table
\ref{tbl:router}. The basic model only copes with 1\% of prefixes; the
ingress one also works on 33\% of the prefixes but not 100\%.  The
egress model is faster than both, and can also symbolically execute
the full router in around 18s.

\headline{Performance comparison to existing tools.}
We seek to understand how \symnet  performance compares to HSA, the most efficient static analysis tool
today. We use the Stanford backbone network data \cite{hsa} and run reachability from an access router to all
core routers with both \symnet  and HSA. The results are given in table \ref{tbl:hsa} and show that 
\symnet is within 50\% of the execution time of HSA, despite its power. In comparison, NOD is reported to
be 20 times slower than HSA \cite{nod} on the same benchmark. 


 
\subsection{Coverage analysis of TCP options code} 
\label{sec:eval:parse}
What properties can Klee 
symbolic execution capture on ASA options parsing C code,
and what can \symnet do on the SEFL model? We stop the tools after one hour.

We ran experiments by creating a TCP packet with a symbolic options field and 
the length field set to a concrete value. We then process this packet with
the options parsing code. Finally, we iterate the options field afterwards,
printing a message if a specific TCP option is present. A list of the 
properties captured by Klee and SymNet is given in Table \ref{tbl:options}.

Table \ref{tbl:klee_options} in \S\ref{sec:examples} shows that Klee
symbolic execution on the C code takes more than 30 mins for only 6B of options.
When options length is less than or equal to six, Klee proves that the parsing code is memory safe,
guaranteed to terminate, and correctly accounts for options with invalid
length. It also shows that the MSS, SackOK and WScale options are allowed
alone or pairwise but not all three simultaneously.  Klees reports that timestamp
and multipath options are not allowed.
Unfortunately, these last results are wrong when we allow full-size
options: the timestamp option is allowed through when the options
field is large enough to contain it. To conclude, Klee testing can
give wrong results when used on small options fields, and can't be run
on large ones. 

\symnet runs the options code in 1s. It cannot capture memory safety
and bounded execution properties of the C code, but ensures the model
itself is memory safe and runs in bounded time by construction. It correctly captures all other properties,
showing that the multipath option is always stripped, the MSS option is always added even if it is not present
in the original packet, and that all allowed options are permitted in any combination. 

\symnet achieves these results by ignoring the ordering of options and by using an abstract representation.
However, ordering matters in the options parsing code: when an option has invalid format (e.g. wrong size), 
the code will replace all remaining bytes with NOPs, thus stripping options stored after the invalid one. 
The \symnet model simply marks \emph{all existing options} in the packet as possibly removed by
setting their corresponding OPT variables to new symbolic values. 



\subsection{Automated testing}

SEFL models can be checked quickly, but they are only useful as
long as they accurately reflect the processing performed by the
baseline code they mimic. As modeling is manual, inadvertent errors
may be introduced.

To catch such bugs, we have developed an automated testing framework
that compares the model to the actual implementation (be it a Click configuration
or a hardware appliance). Our automated tool is similar in principle to ATPG \cite{atpg} and proceeds in 
the following steps:
\begin{enumerate}
\item We run a reachability test over the SEFL model, with a TCP/IP packet with symbolic fields. The output
is a series of paths, where each path places a number of constraints on the header fields of the injected packet. 
\item Pick an unexplored execution path and use Z3 and the path constraints to generate concrete values for
  all the header fields, resulting in a concrete packet $p$.
\item Packet $p$ is injected into the running code (either the Click
  modular router instance or the hardware ASA box). The outputs are captured
  with tcpdump. We use a 1s timeout for tcpdump---if no packets are received, we conclude there is no packet reachable on output.
\item The header values from the captured packet are added as constraints at the end of the symbolic execution path.
Z3 is used to check if the constraints hold; if they do not, an error report is generated. 
\item Repeat from step 2, as long as there are unexplored symbolic execution paths. 
\item Generate random header fields and repeat from step 3.
\end{enumerate}

\begin{figure*}
\begin{minipage}{.28\textwidth}
\includegraphics[width=.8\textwidth,trim=2cm 6cm 7cm 0.9cm,clip=true]{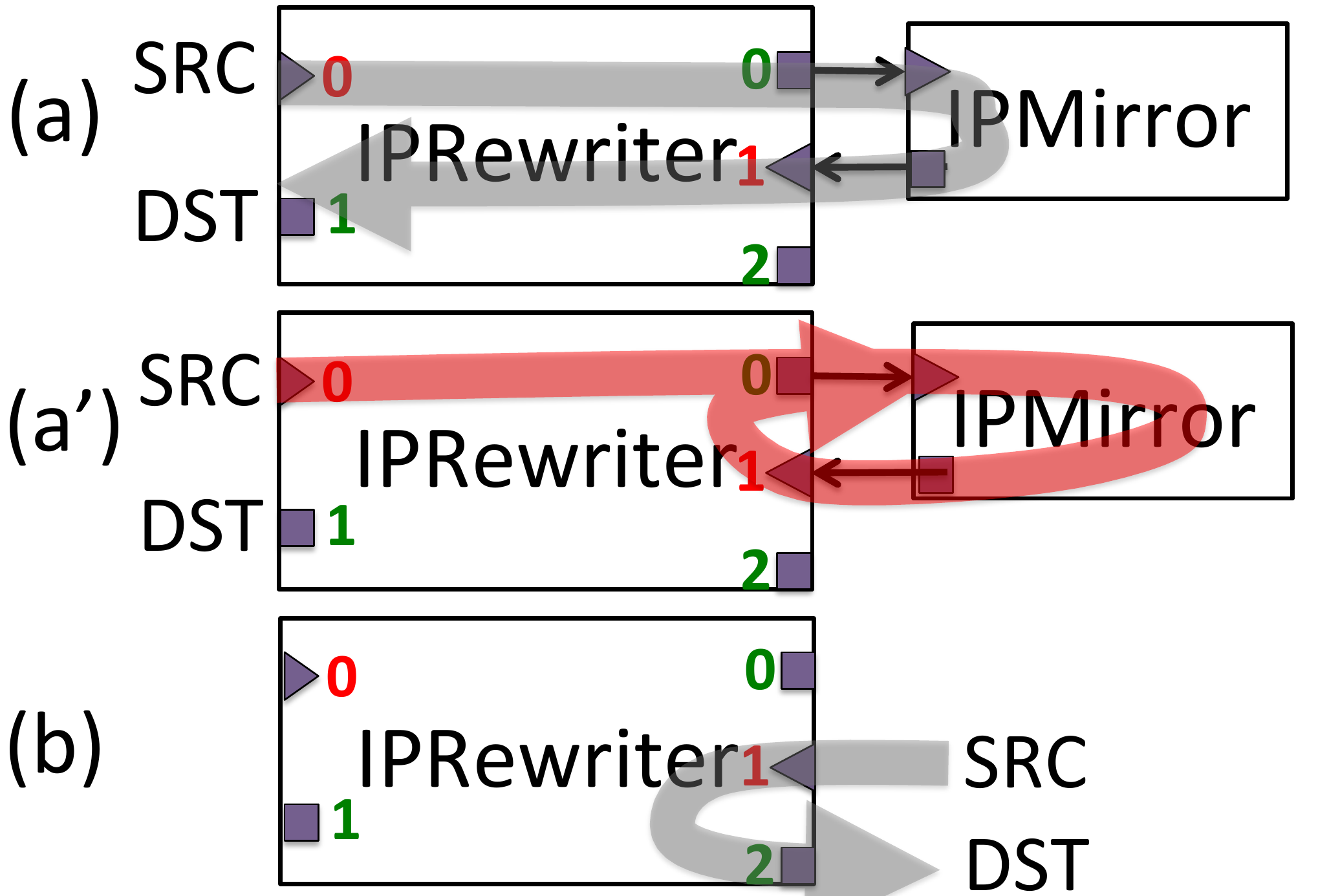}
\vspace{-0.1in}
\caption{Testing a stateful firewall uncovered a cycle.}
\label{fig:iprw}
\end{minipage}
\begin{minipage}{.33\textwidth}
\includegraphics[width=.8\textwidth,trim=2cm 13cm 13cm 2cm]{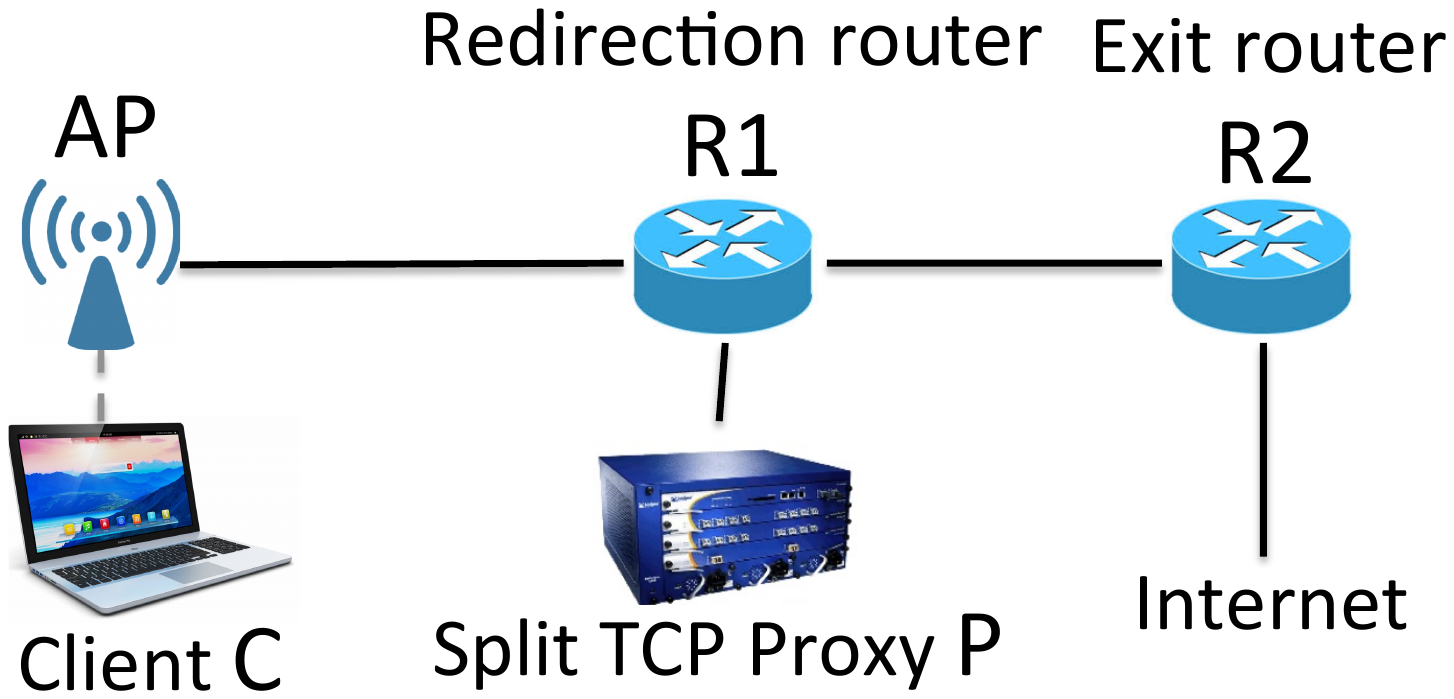}
\vspace{+0.2in}
\caption{Split TCP Deployment, sideband mode \protect\cite{op-middlebox}.}
\label{fig:splittcp}
\end{minipage}
\begin{minipage}{.345\textwidth}
\includegraphics[width=.8\textwidth, trim=2cm 10.3cm 9cm 1cm,clip=true]{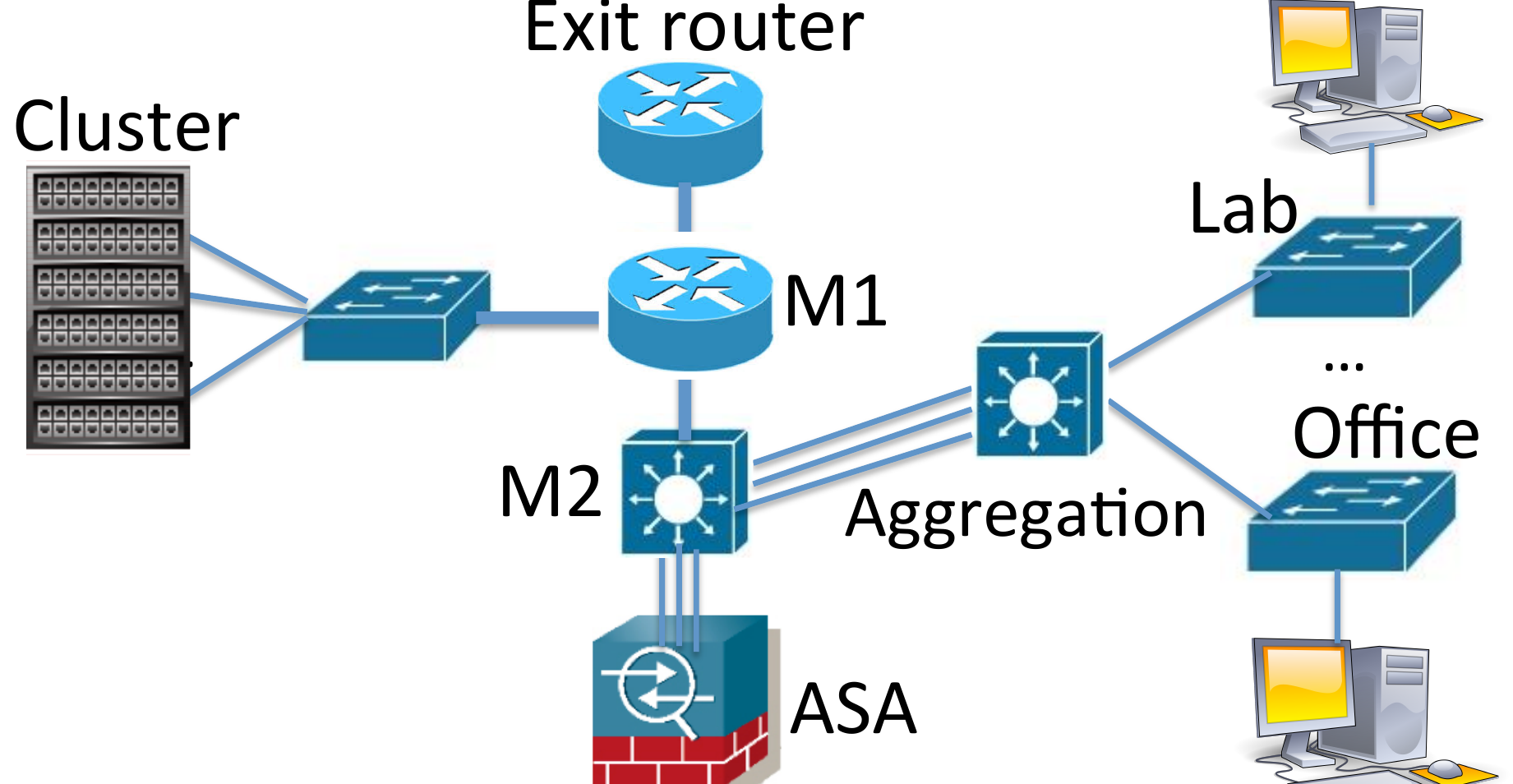}
\vspace{-0.05in}
\caption{CS Department network modeled in SEFL and verified with \symnet}
\label{fig:csnet}
\end{minipage}
\vspace{-0.2in}
\end{figure*}

Our testing procedure first explores all paths resulting from symbolic
execution, and then tests random inputs until the user stops the
testing procedure. We deployed our tool on three machines in our local
testbed: one machine generates packets according to specification,
sends them to the middle machine running the code under test which
sends its output to a third machine where packets are captured.  The
middle machine ran Click modular router configurations for most of our
tests, with one exception: when testing the ASA box we replaced the
middle machine with real ASA hardware configured the same way as our
department's firewall.
We have used testing extensively while writing our models, and it has helped
us catch a series of interesting bugs, discussed below.

\headline{IPMirror.} This Click element mirrors the IP source and destination addresses and transport level ports.
Our model was incomplete: it only mirrored the IP addresses and not ports. 

\headline{DecIPTTL.} This element decrements the IP TTL and drops
packets with TTL 0. The original code decreased the TTL and then
constrained its value to be positive. SymNet reported a single
execution path instead of the two we expected. This was a bug because the TTL is an unsigned value: when
the TTL was 0 decrementing it would result in wrap-around to the
higher possible TTL value, and the packet would never be dropped.  The
obvious fix was to place the constraint $TTL\ge 1$ first, then decrement the TTL.

\headline{HostEtherFilter} only allows packets destined to a MAC address, and we
were wrongly checking the ethertype field. 

\headline{IPClassifier} takes an input packets and forwards them on one of the output ports if the packet matching
the corresponding filter. To test the classifier, we generated multiple instances using different filters and different number of output ports. 
We then took each instance and tested it, one output port at a time, discarding all packets from the other
output ports.  We found a bug where the solver was generating 0 values for several header fields (e.g. port number) and these
where correctly dropped by Click. To solve this issue, we had to constrain our symbolic packet to ensure
IP addresses and port numbers were valid. 

\headline{IPRewriter} is the Click element that allows the implementation of stateful functionality such as NATs and stateful firewalls. 
In our setup, the element acts as a stateful firewall where traffic from the inside network arrives on input port 0 and outside traffic
arrives on input port 1 (Fig. \ref{fig:iprw}). 

To test the rewriter using our unidirectional testing setup we use the setup
shown in Figure \ref{fig:iprw} to testing connections initiated from the inside in setup
 (a) or the outside in setup (b). In setup (a) we model return traffic by bouncing traffic via an IPMirror element.
On output port 0 we should observe a packet with reversed IP addresses and transport ports.

When we ran SymNet we found the loop shown in Fig.\ref{fig:iprw}.(a'): with symbolic packets
it is possible that the source and destination addresses and ports are identical and the 
traffic returned from the IPMirror matches the forward mapping and exits again on output port 0.
The solution was to constrain the source and destination IPs to be different.

\headline{ASA model.} We tested the ASA model against the ASA hardware
extensively, uncovering several bugs in the process. A bug in the VLAN
and ethernet decapsulation code resulted in all packets being dropped
at ASA ingress. Secondly, we found that the ASA correctly allowed
traffic from the higher-priority office VLAN to the lab VLAN, but
wrongly dropped the return traffic. We fixed it by enhancing the ASA
to behave as a stateful firewall for office
to lab traffic.

\subsection{Functional Evaluation}
\label{sec:evaluation_func}

What type of properties
beyond basic IP or layer two reachability can 
\symnet  verify that are useful in practice?
To begin answering this question, we turn to recent work by Le et al. 
that describes operational experiences learned while deploying
a Split TCP middlebox in ten enterprise networks serving thousands of 
users \cite{op-middlebox}. We have modeled in SEFL the 
network topology that Split TCP uses (Figure \ref{fig:splittcp}). The Split TCP Proxy is deployed adjacent to router R3 
which is configured to redirect traffic coming from both directions
to P by rewriting the destination MAC address of the packet. 
Our model faithfully mimicks packet processing along
the whole path, including Ethernet header encapsulation and
decapsulation at each hop, routing and filtering. 
Can \symnet discover issues that appeared in production deployment?

\headline{Asymmetric routing.} We run a reachability check from C to R2,
and at R2 we use IPMirror to send the traffic back to C. \symnet  shows that 
all execution paths from C to R2 and reverse cross via P, thus the setup
is correct.

\headline{MTU issues.} Router R1 is configured to drop all packets
with size large than 1536B. We inject a symbolic packet at C with a symbolic IP 
length field. At R2, the IP length field has a constraint attached: $length<1536$.

Next, we use IP-in-IP tunneling for traffic between R1 and P. This
further reduces the available MTU, and was creating difficult to debug
performance problems in the actual deployment: ping and TCP connection
setup worked fine, but subsequent full MTU traffic from the client was
blackholed because they exceed the MTU after encapsulation
\cite{op-middlebox}. Running reachability on this new setup, the new
constraint applied to length becomes: $length + 20 < 1536$: 
the client MTU must be smaller than 1516B.

\headline{Missing VLAN tagging.} In one setup, P was removing VLAN tags
before processing packets, and was not adding them back before pushing
packets back ro R1. This caused R1 to drop those packets because it 
was expecting VLAN tagging \cite{op-middlebox}. A simple reachability
check quickly highlights this problem: when R1 attempts to remove the
VLAN tagging it finds the wrong EtherType and drops the packet.

\headline{Security Appliance.} In one deployment, R2 acted as a DHCP server
too, and it filtered packets where the Ethernet source address, IP source address
tuple was not in its assigned leases. We modeled the DHCP assignment by using
two metadata variables set by C: \texttt{origIP} and \texttt{origEther}. Both were set by 
the source to have the same symbolic value as the Ethernet 
and IP source address fields in the symbolic packet. R2 filters all packets 
where \texttt{origIP!=pSrc} or \texttt{origEther != EtherSrc}. We ran reachability
again finding all packets were dropped by R2 because the source
MAC was being modified by P and the second constraint didn't hold.

\subsection{Verifying the CS Department Network}

We have generated an accurate model of our department's network using switch MAC tables,
router forwarding tables and a Click configuration mimicking the ASA box. 
The slightly simplified topology is given in
Figure \ref{fig:csnet}: all hosts are connected to local 
switches which connect to an aggregation switch. The 
aggregation switch connects to M2 master switch that connects to a
Cisco ASA box and M1, the CS department router. L2 routing is used in
most of the network: the access switches tag the traffic coming
from the hosts as lab traffic (VLAN 304) or office traffic (VLAN
302). These are carried on trunk links all the way to the ASA box
which is the first IP hop. 
A management VLAN
is also configured with all switches having an interface in it. 
A single server called ``hole'' and located in the cluster has an interface in this VLAN
and is used by our admin for management purposes.
The network has 21 devices with 235 connected network ports. The combined MAC tables have
6000 entries, and there are 400 routing table entries.  

Injecting a purely symbolic packet in the office
takes 42s and results in 3000 paths: results show that the Internet and the labs VLAN are reachable via the ASA
box. If we specialize the packet to have a source address in the office VLAN and a destination address in
the Internet the number of valid paths drops to 50 and execution takes just 10 seconds. Next, we check TCP connectivity
by further specializing the packet (IPProto field is set to 6) and adding an IPMirror element
to the exit router port; this takes 1s and results in 8 valid paths. 

TCP reachability is allowed, but the output shows that TCP options are
tampered with: SACK is disabled for HTTP traffic (OPT5 is set to 0), and MPTCP options
are removed. Our admin had no idea of this behaviour, which was
enabled by the default ASA configuration.

Next, we checked inbound reachability by injecting a purely symbolic packet
at the exit router.  This took 2s and resulted in 221 paths, out of
which 30 were successful. Most paths were ended at the ASA box which
appears to be configured correctly. However, symbolic execution showed
that the management VLAN, with private \emph{192.168.137.0/24} addresses,
was accessible via router M1. A quick check on the live router showed
the behaviour was true, but is it exploitable? Our ISP does not forward 
such traffic towards  to our network. We ran another reachability test
from the cluster and found that any machine from the cluster can access any of the switches,
and verified manually that from the cluster we could indeed telnet into all the switches. 
As all of our students have accounts in the cluster, this is a major security risk.
We have announced it to our admins and they promptly fixed it by updating the static routes at M2.

\section{Related work}
\label{sec:related}

Static network analysis is a well-established topic, 
with many available tools \cite{staticchecking,hsa,anteater,nod,berk}. 
AntEater \cite{anteater} models network boxes as boolean
formulae.
Network Optimized Datalog \cite{nod} is the most complete tool to date and relies on Datalog
both for network models and policy constraints. The work of Panda et al. uses
a model checker to verify networks containing stateful middleboxes
\cite{berk}.  Our NAT model is similar in spirit with their proposal.

In table \ref{tbl:comparison} we provide a qualitative comparison of
the most relevant network verification tools, finding that memory
correctness is a differentiating feature of \symnet. We categorize the
scalability of the tools by analyzing their runtime on
enterprise-sized networks, as reported in the original papers. High
scalability means runtimes of seconds, medium is minutes, while low
means runtimes of hours.  HSA scales very well but it does not capture
many properties. \symnet scales very well on our optimized models; in
general, though, its complexity can quickly run out of control if the
models verified are poorly written or have inherently many branches.

Model independence means a box's model is independent of its location in the network.
NOD is the only tool that doesn't have this property, which makes
modeling very difficult especially in heterogenous networks. 

Finally, we review widely-deployed network functionalities that we should statically verify. \symnet  has the biggest coverage,
however it cannot model packet splitting or coalescing (see our discussion of limitations in \S\ref{sec:limit}).

\begin{table}[t]
\centering
{\mateisize
\begin{tabular}{|p{2.3cm}|p{.65cm}|p{.9cm}|p{.51cm}|p{.51cm}|p{.65cm}|}
\hline
 &HSA\cite{hsa}&AntEater& NOD & Panda & SymNet\\
\hline
Reachability &  \checkmark &  \checkmark &  \checkmark &  \checkmark &  \checkmark \\
Invariants &  \xmark &  \checkmark &  \checkmark &  \checkmark  &  \checkmark \\ 
Header visibility & \xmark  &  \checkmark &  \checkmark &  \checkmark &  \checkmark \\
Memory correctness & \xmark & \xmark & \xmark & \xmark &  \checkmark \\
\hline
Scalability & high & low & med & low & high \\
Model independence & \checkmark & \checkmark & \xmark & \checkmark & \checkmark \\
\hline
IP router &  \checkmark & \checkmark & \checkmark & \checkmark &  \checkmark \\
Dynamic tunneling &  \xmark & \xmark & \xmark & \xmark &  \checkmark \\
TCP options &  \xmark & \xmark & \checkmark & \xmark &  \checkmark \\
Dynamic NATs &  \xmark& \xmark & \checkmark & \checkmark &  \checkmark \\
Encryption &  \xmark & \xmark & \xmark & \xmark &  \checkmark \\
TCP Segment splitting &  \xmark & \xmark & \xmark & \xmark &  \xmark \\
IP Fragmentation  &  \xmark & \xmark & \xmark & \xmark &  \xmark \\
\hline
\end{tabular}
}
\vspace{-0.1in}
\caption{\symnet  vs. other network verification tools.}
\vspace{-0.2in}
\label{tbl:comparison}
\end{table}

\headline{Symbolic execution.} We are not the first to propose using symbolic execution to analyze 
networks. Dobrescu et al. \cite{symbex-dobrescu} used
symbolic execution to check selected Click elements' source code for 
bugs, aiming to proove crash-freedom and bounded execution.
We have shown that using C as modelling language does not scale,
and have proposed SEFL and \symnet as scalable alternatives.

\headline{Online verification.} Veriflow \cite{veriflow} and NetPlumber \cite{netplumber} aim to
perform live validation of all network configuration changes. They
work underneath an SDN controller and verify all state updates. NICE
uses symbolic model checking to verify the correctness of Openflow
programs \cite{nice}. 
More recently,
Armstrong \cite{symbex-vyas} use Klee on middlebox models written
in C to guide the generation of test packets for networks. \symnet  is
orthogonal to these works.

NetKAT\cite{netkat} and Frenetic \cite{frenetic} are novel specification languages
optimized for specifying OpenFlow-like rules in networks. SEFL is strictly more general
as it can model middlebox behaviours too, not just layer two behaviour. 


\section{Limitations}
\label{sec:limit}
Using \symnet for network analysis is powerful but has a few notable limitations which we 
discuss here. 
Packet processing is sequential, and the network boxes only do
processing when they receive packets.  This problem is not unique to
\symnet: it is a general limitation of applying symbolic execution
to network analysis.  Parallel processing is akin to
symbolic execution of multi-threaded programs, and is significantly harder
because it must check all possible interleavings
of the different threads \cite{parallel-symbex}.

A single packet is active in the network at any one time. This implies
that processing that works across multiple packets, such as TCP
segment splitting or coalescing, cannot be modelled. 
This only limits the number of in-flight packets,
not the number of total packets: our TCP sender model works with a
window of one packet because of this limitation, but any
number of round-trip times can be simulated.


\section{Conclusions}
\label{sec:conclusions}





\vspace{-0.05in}Symbolic execution is a powerful tool for network verification,
but applying it to production networks is challenging.
To allow scalable network symbolic execution we have proposed SEFL, a novel,
minimalist, imperative language tailored by design for network
symbolic execution. We have built 
\symnet, a fast symbolic execution tool for SEFL code. 

To understand the expressiveness of SEFL, we have modeled many
networking devices ranging from switches and routers to middleboxes that
keep flow state and parse TCP options. We have also modeled 
a large subset of the elements of the Click modular router, which allows
us to verify Click configurations out-of-the box. 
Our evaluation shows that our models have near-optimal branching factors
per box and that \symnet  seamlessly scales to large networks. 

Finally, we have used \symnet  to capture a number of middlebox
behaviours described in the literature and in our own department's network.
Our experience shows that \symnet catches many interesting network 
properties and is fast.

\balance
\bibliographystyle{habbrv}
\bibliography{biblio}
\end{document}